\documentclass[runningheads]{llncs}
\usepackage[T1]{fontenc}
\usepackage{graphicx}
\usepackage{xspace}
\usepackage{mdframed}
\usepackage{amsmath,amssymb}
\usepackage{cite,enumitem}
\usepackage{hyperref}

\usepackage[misc,geometry]{ifsym}
\usepackage{listings}
\usepackage{pgfplots}
\pgfplotsset{width=5.75cm, compat=1.17}

\newcommand{\mcsat}{MCSat\xspace}
\newcommand{\Fp}{\ensuremath{\mathbb{F}_p}}

\newcommand{\FpX}{\ensuremath{\Fp[X]}}
\newcommand{\poly}[1]{\ensuremath{\mathsf{poly}(#1)}}

\newcommand{\yices}{\textsc{Yices2}\xspace}
\newcommand{\cvc}{\textsc{cvc5}\xspace}
\newcommand{\libpoly}{\textsc{LibPoly}\xspace}
\newcommand{\smtlib}{SMT-LIB~2\xspace}

\def\orcidID#1{\href{http://orcid.org/#1}{\raisebox{-1.25pt}{\includegraphics{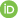}}}}

\begin{document}
	\title{\mcsat-based Finite Field Reasoning in the \yices SMT Solver}
	\titlerunning{\mcsat-based Finite Field Reasoning in the \yices SMT Solver}
	\author{Thomas Hader\inst{1}\textsuperscript{(\Letter)} \and
		Daniela Kaufmann\inst{1}\orcidID{0000-0002-5645-0292} \and
		Ahmed Irfan\inst{2}\orcidID{0000-0001-7791-9021} \and \newline
		St\'ephane Graham-Lengrand\inst{2}\orcidID{0000-0002-2112-7284} \and
		Laura Kov\'acs\inst{1}\orcidID{0000-0002-8299-2714}
	}
	\authorrunning{Hader et al.}
	\institute{
		TU Wien, Vienna, Austria \\
		\email{\{thomas.hader,daniela.kaufmann,laura.kovacs\}@tuwien.ac.at}
		\and
		SRI International, Menlo Park, CA, USA \\
		\email{\{ahmed.irfan,stephane.graham-lengrand\}@sri.com}
	}
	\maketitle              %
	\setcounter{footnote}{0} 
	\begin{abstract}
		This system description introduces an enhancement to the \yices SMT solver, enabling it to reason over non-linear polynomial systems over finite fields. 
        Our reasoning approach fits into the model-constructing satisfiability (\mcsat) framework
        and is based on zero decomposition techniques, which find finite basis explanations for theory conflicts over finite fields.
        As the \mcsat solver within \yices can support (and combine) several theories via theory plugins,
        we implemented our reasoning approach as a new plugin for finite fields
        and extended \yices's frontend to parse finite field problems,
        making our implementation the first \mcsat-based reasoning engine for finite fields.
        We present its evaluation on finite field benchmarks, comparing it against \cvc.
        Additionally, our work leverages the modular architecture of the \mcsat solver in \yices
        to provide a foundation for the rapid implementation of further reasoning techniques for this theory.
	\end{abstract}

	\section{Introduction}
 Satisfiability Modulo Theories (SMT) solving plays a crucial role in
 automated reasoning  as it combines the power of Boolean
 satisfiability (SAT) with various mathematical background
 theories~\cite{smt}. This connection enables the automated
 verification and synthesis of
 systems~\cite{DBLP:journals/cacm/MouraB11} that require reasoning in
 more expressive logical theories, for example real/integer arithmetic.

 State-of-the-art SMT solvers employ a combination of Boolean level reasoning and theory-specific algorithms.
 This is achieved either through the use of the CDCL(T) paradigm~\cite{DBLP:conf/lpar/NieuwenhuisOT04} or the model-constructing satisfiability (\mcsat) approach~\cite{mcsat,mcsat2}.
 The \mcsat algorithm lifts the Boolean-level CDCL algorithm to the theory level,
 while keeping the search theory independent. %
 This approach is particularly effective for handling complex arithmetic theories.
 For instance, \yices~\cite{yices} uses the \mcsat approach to handle non-linear arithmetic constraints.
 
 Finite fields offer an ideal framework for modeling bounded machine arithmetic, particularly relevant in the context of contemporary cryptosystems utilized in system security and post-quantum cryptography. Current methodologies, for instance, develop private and secure systems using zero-knowledge (ZK) proofs~\cite{GoldwaserMicaliRackoff-SIAM89} or authenticate blockchain technologies like smart contracts~\cite{Szabo2018SmartC}. 
Verifying applications in such areas require efficient SMT solvers that support reasoning over finite field arithmetic, e.g., verification of a compiler for ZK proofs~\cite{ozdemir2023cavzkproofs}.

\paragraph{Related Work.}
Currently, the related work on SMT solving in finite field arithmetic is still rather limited. 
Our own theoretical work~\cite{hader2023lpar} on \mcsat approaches
based on finding zero decompositions comes with a proof-of-concept
implementation that %
facilitates only a very fundamental \mcsat algorithm, has only limited support of Boolean propagation, and is unable to parse \smtlib~\cite{SMTLIB}.	 

The only other SMT solver that we are aware of being capable of reasoning over finite fields is \cvc~\cite{ozdemir2023cav,DBLP:conf/tacas/BarbosaBBKLMMMN22}, which uses a classical CDCL(T) approach. 
As a theory engine, Gröbner bases~\cite{Buchberger} reasoning over a set of polynomial equalities is applied. 
If the derived Gröbner basis contains the constant $1$, then the system is unsatisfiable and a conflict core for the CDCL(T) search can be found. 
Otherwise, a guided enumeration of all possible solutions is performed to search for a model.

Note that both approaches~\cite{hader2023lpar,ozdemir2023cav} use complementary techniques. 
On the one hand, Gröbner bases are highly engineered to find conflicts in the polynomial input, which tends to help for unsatisfiable instances~\cite{ozdemir2023cav}. 
On the other hand, a model constructing approach tends to be fast whenever there is a solution, especially when there is a high number of models~\cite{hader2023lpar}.

We further note that at the moment our implementation in \yices, as well as \cvc, is restricted to finite fields where the order (i.e. size) is prime. This limitation is sufficient for many applications in cryptosystems and ZK proofs.

Besides using dedicated finite field solvers, problems over prime fields can be encoded in integer arithmetic using the modulo operator.
Further, since terms are bounded, encoding as bit-vectors for subsequent bit-blasting is possible.
However, prior experiments have shown that those encodings perform poorly on existing solvers~\cite{ozdemir2023cav}.

	\paragraph{Contributions.}
	We present an integration of the theory of non-linear finite field arithmetic in the \yices SMT solver~\cite{yices}, enabling it to reason over finite field problems. This includes the following contributions which we will further explain throughout the rest of this paper:
	\begin{itemize}[wide=0em,leftmargin=1em]
		\item Adding the parsing of finite field problems to
                  the \yices \smtlib front-end (Section~\ref{sec:usability}).
		\item Representing finite field polynomials as terms
                  in \yices and implementing features regarding such
                  polynomials in the \libpoly
                  library~\cite{jovanovic2017smt}, which is the
                  library used for polynomials in \yices (Section~\ref{sec:implementation}).
		\item Implementing and evaluating an \mcsat theory back-end for
                  finite field reasoning, using existing concepts from
                  non-linear real and bit-vector solving from  \yices (Section~\ref{sec:eval}).
	\end{itemize}
	
	 To the best of our knowledge, our work is currently the only finite field instantiation of \mcsat. While our initial theory reasoning approach follows closely the explanation generation procedure of our previous work~\cite{hader2023lpar}, our implementation allows easy drop-in of an improved explanation procedure in the future.

         \section{Preliminaries}
  In mathematics, a finite field is a field that contains a finite number of elements. A finite field $\Fp$ of prime order $p$ can roughly be seen as the representation of the integers modulo the prime $p$. 
  We refer to~\cite{hader2023lpar,ozdemir2023cav} for details on the theory and representation of finite fields. 
Since there is no inherent order on finite fields, polynomial constraints are either equalities $p = 0$ or disequalities $p \neq 0$ for a finite field polynomial $p$.

 For SMT solving in finite fields,  we are interested in the following problem:
	\begin{mdframed}
		Given a finite field $\mathbb{F}_p$, where $ p
		$ is a prime number, let $X= x_1,\ldots x_n$, let $F$ be a set of polynomial
		constraints in $\FpX$ and $\mathcal{F}$ a formula following the logical structure:
		\[\mathcal{F} \quad= \quad \bigwedge_{C\subseteq F} \bigvee_{f\in C} f  \quad=  \quad \bigwedge_{C\subseteq F} \bigvee_{f\in C} \poly{f} \vartriangleright 0 \quad \text{with}   \vartriangleright \in \{=, \neq\}. \]
		\emph{SMT solving over finite fields:}  Does an assignment $ \nu : \{x_1,\dots,x_n\} \rightarrow \Fp^n $ exist that satisfies $\mathcal{F}$?
		\end{mdframed}
	
  For example, the formula $\mathcal{F}_1 = (x-1=0 \lor y-1=0) \land (xy-1 =0)$ is satisfied by the assignment $x \mapsto 1, y\mapsto 1$ in $\mathbb{F}_3$; whereas the formula $\mathcal{F}_2 = (x-1=0 \lor y-1=0) \land (xy-1 =0) \land (x-2 = 0)$ is unsatisfiable in $\mathbb{F}_3$.

	\paragraph{Yices2 and \mcsat.} 

        \yices contains two main solvers, one based on the traditional
        CDCL(T) approach~\cite{DBLP:conf/lpar/NieuwenhuisOT04} and one
        based on the \mcsat approach~\cite{nlsat,mcsat}.  \yices's
        common API and front-ends can automatically select which
        solver to use at runtime, depending on an \smtlib logic. In
        particular, when non-linear real or integer arithmetic
        constraints are present \yices selects the \mcsat solver.  The
        \mcsat solver in \yices currently supports the theories of
        non-linear real arithmetic (QF\_NRA)~\cite{nlsat} and integer
        arithmetic (QF\_NIA)~\cite{DBLP:conf/vmcai/Jovanovic17},
        bit-vectors (QF\_BV)~\cite{mcsat_bv}, equality and
        uninterpreted functions (QF\_EUF), arrays~\cite{mcsat_arrays},
        and combinations thereof.

        In contrast to CDCL(T) that \emph{complements} CDCL with
        theory reasoning, \mcsat applies CDCL-like mechanisms to
        \emph{perform} theory reasoning. Specifically, it explicitly
        and incrementally constructs models with first-order variable
        assignments---maintained in a \emph{trail}---while maintaining
        the invariant that none of the constraints evaluate to false.
        \mcsat decides upon such assignments when there is choice,
        it can propagate them when there is not, and it backtracks upon
        conflict. The lemmas learned upon backtracking are based on theory-specific
        explanations of conflicts and propagations. This theory-specific reasoning is
        implemented through \emph{plugins} that provide interfaces to make
        decisions, perform propagations, and produce explanations.

	\section{Usability of SMT Solving in Finite Fields}\label{sec:usability}
	Support for finite field reasoning in \yices is available on the master branch%
	\footnote{Available at \url{https://github.com/SRI-CSL/yices2}}%
	and will be included in the next release (2.7).
	The theory of finite fields can be accessed using a not-yet official extension of the  \smtlib language (.smt2).

	\paragraph{\smtlib Parsing.} Extending the parser to handle finite field problems was our first extension to \yices.
	Currently, polynomials over finite fields are no official theory in \smtlib~\cite{SMTLIB}.
	However, when implementing finite field support in \cvc~\cite{DBLP:conf/tacas/BarbosaBBKLMMMN22}, an extension was proposed in~\cite{ozdemir2023cav}.
	In the interest of keeping inputs and benchmarks comparable, we aimed at a compatible implementation.
	Standardization efforts to create an official \smtlib theory for finite field arithmetic are currently ongoing.
	
	In the \smtlib extension, the theory of (quantifier-free) non-linear finite field arithmetic is denoted as \lstinline|QF_FFA|. Elements can be defined using the sort \lstinline|FiniteField|.
	The sort is indexed by the order of the finite field, which is required to be a prime number. For instance \lstinline|(_ FiniteField 3)| 
	 defines the finite field of order $3$. 
	 Constants are indexed with the field order to indicate which
         finite field they belong in, e.g., \lstinline|(_ ff2 3)|. Note that the integer following \lstinline|ff| is
         interpreted modulo the field order. As a short-cut to avoid
         rewriting the field order over and over again, the \lstinline|as| keyword can be used to interpret the constant in the correct field type: \lstinline|(as ff2 FF3)| for a defined finite field sort \lstinline|FF3|.
	To specify the finite field operations \lstinline|ff.mul| and \lstinline|ff.add| are available for multiplication and addition of finite field values, respectively. Both support an arbitrary number of operators. Atoms with finite field terms can be \lstinline|=| with its respective meaning.
	For example, an encoding of $\mathcal{F}_1$ can be seen in Fig.~\ref{smtlib2-code}.
	\begin{figure}
	\vspace{-1em}
	\begin{lstlisting}[basicstyle=\footnotesize\ttfamily]
(set-logic QF_FFA)
(define-sort FF3 () (_ FiniteField 3))
(declare-fun x () FF3)
(declare-fun y () FF3)
(assert (and
	(or (= (ff.add x (as ff-1 FF3)) (as ff0 FF3))
	    (= (ff.add y (as ff-1 FF3)) (as ff0 FF3)))
	(= (ff.mul x y) (as ff-1 FF3))))
(check-sat)
	\end{lstlisting}
	\vspace{-1em}
	\caption{Example for an \smtlib encoding of a finite field problem $\mathcal{F}_1$.}
	\label{smtlib2-code}
	\vspace{-2em}
	\end{figure}

	\section{Implementation Details}\label{sec:implementation}

	\paragraph{The Implementation of \mcsat in \yices.} 
        The \mcsat solver in \yices supports multiple theories
        via a notion of theory \emph{plugin} that builds upon an earlier architecture~\cite{mcsat2}.
        An \mcsat theory plugin in \yices implements a number of functionalities that are given to the main \mcsat solver as function pointers.
        The main \mcsat loop calls these functions for theory-specific operations such as deciding or propagating the value of variables or getting explanation lemmas, or upon certain events such as the creation of new terms and lemmas.
        In return, a theory plugin can access theory-generic mechanisms for, e.g., inspecting the \mcsat trail, creating variables and requesting to be notified of certain events like variable assignments, as well as raising conflicts.
        A theory plugin is not required to implement mechanisms for propagating theory assignments and explaining them, but for the current theories in \yices, %
        such propagations have provided noticeable speed-ups (see, e.g.,~\cite{mcsat_bv}).

	\paragraph{The Finite Field \mcsat Plugin.}
	Before handling constraints in the finite field \mcsat plugin, the input assertions are represented as polynomial constraints. Limited preprocessing (e.g., constant propagation) is performed at this step.
	Internally, the plugin only handles polynomial equalities and disequalities.
	The implementation of the finite field plugin follows an approach similar to the plugin for non-linear arithmetic~\cite{DBLP:conf/vmcai/Jovanovic17}.

	Using the \mcsat trail,
        the finite field plugin reads which constraints must be fulfilled at any given time
        (as decided or deduced by the Boolean plugin)
        and tracks the assignment of values to polynomial variables.
        It also tracks, for each polynomial variable,
        the \emph{set of feasible values} that the variable can be assigned
        without any of the polynomial constraints evaluating to false:
        Using watch lists, it detects when any of the constraints becomes \emph{unit},
        i.e. when all of its variables but one have been assigned values.
        Upon such detection,
        it computes how the constraint restricts the set of feasible values for the last remaining variable,
        using regular univariate polynomial factorization.
        When that set becomes empty, the plugin reports a theory conflict to the main \mcsat engine.
	Given a conflict core and the current assignment, the \emph{explanation procedure} in the plugin generates a (globally valid) lemma that explains the conflict in that it excludes a class of assignments (including the current one) that all violate the conflict core.
        The \mcsat engine performs conflict analysis using theory explanations and Boolean resolutions,
        and either backtracks if it can or concludes unsatisfiability.
	On the other hand, the instance is satisfiable once all variables are assigned a value.
	
	\paragraph{Finite Field Explanations.}
	In our earlier work~\cite{hader2023lpar}, we presented an explanation procedure for finite fields.
	This approach employs subresultant regular subchains (SRS)~\cite{wang2001elimination} between conflicting polynomials to provide new polynomial constraints that can be propagated. 
In a nutshell, SRS can be used to construct a generalized greatest common divisor (GCD) of polynomials that takes into account the current partial variable assignment on the trail.
	The computed GCD is utilized in a zero decomposition procedure to reduce the degree of the conflicting polynomials until we can learn a polynomial constraint that excludes the current partial assignment. This constraint is added as an explanation clause to resolve the conflict.	%
	We implemented the procedure of~\cite{hader2023lpar} in the current version of \yices using \libpoly.
	However, it is important to note that there are other solving techniques for polynomial systems over finite field that could potentially be utilized to develop an explanation method suitable for an \mcsat-based search.
	Furthermore, it is still an open question how different techniques perform in an \mcsat environment.
        That is why we have kept the explanation procedure encapsulated in our implementation, allowing for easy extension in order to support development and evaluation of future explanation procedures.
	
	\section{Evaluation}\label{sec:eval}
	Since finite field solving is a rather new endeavor in the world of SMT, no extensive set of \smtlib benchmarks exists yet. 
	For the evaluation we have selected the benchmark sets presented in the papers describing the theory behind the implementation of \yices and \cvc:
	\begin{enumerate}[label=(\roman*),wide=0em,leftmargin=0em]
	\item The polynomial sets from our prior work~\cite{hader2023lpar}, consisting of 325 instances.
	These benchmarks primarily contains finite fields up to
        order $211$, using two classes of polynomial systems: \emph{randomly generated} as well as \emph{crafted} systems. 
        The crafted benchmarks are product of (mostly) degree-1 polynomials.
	\item Benchmarks generated using ZK proof compilers presented in~\cite{ozdemir2023cav}.
	Besides polynomial equations, these 1602 benchmark instances also contain Boolean structure. The field order varies form small ($11$) up to vast (more than $2^{255}$).
	\end{enumerate}

	\paragraph{Experimental Setup.}
	Our experiments were run on an AMD EPYC 7502 CPU with a timeout of $300$ seconds per benchmark instance.
	We compared our implementation of \yices against \cvc version 1.1.1, which is the latest released version at the time of writing. We are not aware of any further SMT solvers supporting the theory of non-linear finite fields to be included in the comparison.

	\paragraph{Experimental Results.}
	The performance comparison between the two solvers on the first benchmark set can be seen in Fig.~\ref{fig:scatter} and Fig.~\ref{fig:cactus} (left).
	It is clear to see that the random instances are harder to solve than the crafted instances (which have significantly more variables).
	We believe that this is due to the lack of internal structure in random polynomials.
	This makes symbolic handling of those polynomial systems hard, both for Gröbner basis computation (in \cvc) as well as SRS computation (in \yices).

\begin{figure}
	\vspace{-1em}
	\begin{tikzpicture}
		\begin{loglogaxis}[
			title={Random benchmarks},
			xlabel={\yices},
			ylabel={\cvc},
			grid=major,
			xmin=0.001, xmax=300,
			ymin=0.001, ymax=300,
			xtick={0.01,0.1,1,10,100},
			ytick={0.01,0.1,1,10,100},
			clip=true,
			enlargelimits=false
			]
			\draw (axis cs:0.001,0.001) -- (axis cs:300,300);
			\addplot [
				blue,
				scatter,
				only marks,
				point meta=explicit symbolic,
				scatter/classes={
					sat={mark=o},
					unsat={mark=x},
					unknown={mark=x,draw=none}
				},
			] table [meta=result] {data/testdata_i_3_8_8.csv};
			\addplot [
				blue,
				scatter,
				only marks,
				point meta=explicit symbolic,
				scatter/classes={
					sat={mark=o},
					unsat={mark=x},
					unknown={mark=x,draw=none}
				},
			] table [meta=result] {data/testdata_i_3_16_16.csv};
			\addplot [
				green,
				scatter,
				only marks,
				point meta=explicit symbolic,
				scatter/classes={
					sat={mark=o},
					unsat={mark=x},
					unknown={mark=x,draw=none}
				},
			] table [meta=result] {data/testdata_i_13_8_4.csv};
			\addplot [
				green,
				scatter,
				only marks,
				point meta=explicit symbolic,
				scatter/classes={
					sat={mark=o},
					unsat={mark=x},
					unknown={mark=x,draw=none}
				},
			] table [meta=result] {data/testdata_i_13_8_8.csv};
			\addplot [
				orange,
				scatter,
				only marks,
				point meta=explicit symbolic,
				scatter/classes={
					sat={mark=o},
					unsat={mark=x},
					unknown={mark=x,draw=none}
				},
			] table [meta=result] {data/testdata_i_211_8_4.csv};
			\addplot [
				orange,
				scatter,
				only marks,
				point meta=explicit symbolic,
				scatter/classes={
					sat={mark=o},
					unsat={mark=x},
					unknown={mark=x,draw=none}
				},
			] table  [meta=result] {data/testdata_i_211_8_16.csv};
		\end{loglogaxis}
	\end{tikzpicture}
	\hspace{1em}
	\begin{tikzpicture}
		\begin{loglogaxis}[
			title={Crafted benchmarks},
			xlabel={\yices},
			ylabel={\cvc},
			grid=major,
			xmin=0.001, xmax=300,
			ymin=0.001, ymax=300,
			xtick={0.01,0.1,1,10,100},
			ytick={0.01,0.1,1,10,100},
			clip=true,
			enlargelimits=false
			]
			\draw (axis cs:0.001,0.001) -- (axis cs:300,300);
			\addplot [
				blue,
				scatter,
				only marks,
				point meta=explicit symbolic,
				scatter/classes={
					sat={mark=o},
					unsat={mark=x},
					unknown={mark=x,draw=none}
				},
			] table [meta=result] {data/testdata_r_3_32_32.csv};
			\addplot [
				blue,
				scatter,
				only marks,
				point meta=explicit symbolic,
				scatter/classes={
					sat={mark=o},
					unsat={mark=x},
					unknown={mark=x,draw=none}
				},
			] table [meta=result] {data/testdata_r_3_64_64.csv};
			\addplot [
				green,
				scatter,
				only marks,
				point meta=explicit symbolic,
				scatter/classes={
					sat={mark=o},
					unsat={mark=x},
					unknown={mark=x,draw=none}
				},
			] table [meta=result] {data/testdata_r_13_32_8.csv};
			\addplot [
				green,
				scatter,
				only marks,
				point meta=explicit symbolic,
				scatter/classes={
					sat={mark=o},
					unsat={mark=x},
					unknown={mark=x,draw=none}
				},
			] table [meta=result] {data/testdata_r_13_32_16.csv};
			\addplot [
				orange,
				scatter,
				only marks,
				point meta=explicit symbolic,
				scatter/classes={
					sat={mark=o},
					unsat={mark=x},
					unknown={mark=x,draw=none}
				},
			] table [meta=result] {data/testdata_r_211_8_6.csv};
			\addplot [
				orange,
				scatter,
				only marks,
				point meta=explicit symbolic,
				scatter/classes={
					sat={mark=o},
					unsat={mark=x},
					unknown={mark=x,draw=none}
				},
			] table  [meta=result] {data/testdata_r_211_16_8.csv};
		\end{loglogaxis}
	\end{tikzpicture}
	\vspace{-1em}
	\caption{Runtime comparison for benchmarks from~\cite{hader2023lpar} (in seconds, timeout 300s)\\
	result: sat \textsf{o}, unsat \textsf{x}; finite field order: {\color{blue}3 (blue)}, {\color{green}13 (green)}, and {\color{orange}211 (orange)} }
	\label{fig:scatter}
	\vspace{-1em}
\end{figure}
	
	Note that \cvc performs most symbolic computation upfront (when generating the Gröbner basis) and enumerates potential solutions in a second step (using auxiliary Gröbner basis calls). The \mcsat approach in \yices, on the other hand, interleaves model generation and symbolic computation during the search.
	This tends to be an advantage for harder polynomial systems especially together with small finite field orders. When the finite field order increases, this advantage seems to vanish.
	For the crafted polynomial benchmarks, \yices tends to be faster. We believe that this is due to the fact that the polynomials tend to be large (in the number of monomials), but rather easy to solve.
	Generating a full Gröbner basis upfront might add significant overhead.
	
	For the second benchmark set, many instances can be solved by both solvers immediately (c.f.\ Fig.~\ref{fig:cactus} right).
	We believe that those instances can be solved without extensive finite field reasoning, as the benchmark set contains Boolean structure.
	This enables both solvers to successfully solve benchmarks even with vast field orders.
	However, once extensive algebraic reasoning is required in finite fields of vast order (the majority of the benchmarks), the purely symbolic approach of \cvc in proving unsatisfiability seems to be advantageous.
	An \mcsat approach requires to pick actual values in a gigantic search space, thus especially strong lemmas need to be learned in order to prune the search space efficiently.
	Improving the explanation procedure is part of our future work.
	
	\begin{figure}
		\vspace{-0.5em}
		\begin{tikzpicture}
			\begin{axis}[
				axis lines=left,
				xmin=0, xmax=325,
				ymin=0.006, ymax=300,
				minor x tick num=1,
				minor y tick num=5,
				ymode=log,
				ytick={0.01,0.1,1,10,100},
				grid=major,
				ylabel={time (s)},
				legend style={at={(0.02,0.9)}, anchor=west}
				]
				\draw[dashed] (0,300) -- (325,300); 
				\draw[dashed] (325,0.006) -- (325,300); 
				\addplot[blue] table {data/ffsat.yices.cactus.csv};
				\addplot[red] table {data/ffsat.cvc5.cactus.csv};
			\end{axis}
		\end{tikzpicture}
		\hspace{1em}
		\begin{tikzpicture}
			\begin{axis}[
				axis lines=left,
				xmin=0, xmax=1602,
				ymin=0.006, ymax=300,
				minor x tick num=1,
				minor y tick num=5,
				grid=major, 
				ymode=log,
				ytick={0.01,0.1,1,10,100},
				ylabel={time (s)},
				legend style={at={(0.02,0.9)}, anchor=west}
				]
				\draw[dashed] (0,300) -- (1602,300); 
				\draw[dashed] (1602,0.006) -- (1602,300); 
				\addplot[blue] table {data/circ.yices.cactus.csv};
				\addplot[red] table {data/circ.cvc5.cactus.csv};
			\end{axis}
		\end{tikzpicture}
		\vspace{-1.6em}
		\caption{Instances solved over time (timeout 300s) by \yices ({\color{blue} blue}) and \cvc ({\color{red} red}) from~\cite{hader2023lpar} (left) and~\cite{ozdemir2023cav}~(right).}
		\label{fig:cactus}
		\vspace{-1em}
	\end{figure}

	\section{Summary and Outlook}
	In this system description we have presented the first implementation of an \mcsat-based decision procedure for non-linear finite field polynomials. 
	We have shown that \mcsat is a feasible way of solving SMT instances over finite fields and it compares well with SMT approaches using Gröbner bases for many instances.
	
	The presented tool implementation is well suited for future experiments and the rapid development of more advanced explanation procedures that will eliminate the current bottlenecks with regard to large finite fields.

	\begin{credits}
		\subsubsection{\ackname} This work was conducted during the first author's stay at SRI International. 
		We acknowledge funding from the ERC Consolidator Grant ARTIST 101002685, the
		TU Wien SecInt Doctoral College, the FWF  grants SFB
                10.55776/F8504 and ESPRIT 10.55776/ESP666), the
		WWTF ICT22-007 project ForSmart, the NSF award
                CCRI-2016597, 
                the Amazon Research Award 2024 QuAT, and from SRI Internal Research And Development funds.
                Any opinions, findings and conclusions or recommendations
                expressed in this material are those of the author(s) and do not necessarily
                reflect the views of the US Government or NSF.
		\subsubsection{\discintname}
		The authors have no competing interests to declare that are relevant to the content of this article.
	\end{credits}
	\bibliographystyle{splncs04}
	\bibliography{ffsat}

\begin{thebibliography}{10}
\providecommand{\url}[1]{\texttt{#1}}
\providecommand{\urlprefix}{URL }
\providecommand{\doi}[1]{https://doi.org/#1}

\bibitem{DBLP:conf/tacas/BarbosaBBKLMMMN22}
Barbosa, H., Barrett, C.W., Brain, M., Kremer, G., Lachnitt, H., Mann, M.,
  Mohamed, A., Mohamed, M., Niemetz, A., N{\"{o}}tzli, A., Ozdemir, A.,
  Preiner, M., Reynolds, A., Sheng, Y., Tinelli, C., Zohar, Y.: cvc5: {A}
  versatile and industrial-strength {SMT} solver. In: Fisman, D., Rosu, G.
  (eds.) Intl. Conf. on Tools and Algorithms for the Construction and Analysis
  of Systems {(TACAS)} , Part {I}. LNCS, vol. 13243, pp. 415--442. Springer
  (2022). \doi{10.1007/978-3-030-99524-9\_24},
  \url{https://doi.org/10.1007/978-3-030-99524-9\_24}

\bibitem{SMTLIB}
Barrett, C., Fontaine, P., Tinelli, C.: {The Satisfiability Modulo Theories
  Library (SMT-LIB)}. {\tt www.SMT-LIB.org} (2016)

\bibitem{smt}
Barrett, C.W., Sebastiani, R., Seshia, S.A., Tinelli, C.: Satisfiability modulo
  theories. In: Biere, A., Heule, M., van Maaren, H., Walsh, T. (eds.) Handbook
  of Satisfiability - Second Edition, Frontiers in Artificial Intelligence and
  Applications, vol.~336, pp. 1267--1329. {IOS} Press (2021).
  \doi{10.3233/FAIA201017}, \url{https://doi.org/10.3233/FAIA201017}

\bibitem{Buchberger}
Buchberger, B.: {Bruno {Buchberger's PhD} Thesis 1965: An Algorithm for Finding
  the Basis Elements of the Residue Class Ring of a Zero Dimensional Polynomial
  Ideal}. Journal of Symbolic Computation  \textbf{41}(3-4),  475--511 (2006).
  \doi{10.1016/J.JSC.2005.09.007},
  \url{https://doi.org/10.1016/j.jsc.2005.09.007}

\bibitem{yices}
Dutertre, B.: Yices 2.2. In: Biere, A., Bloem, R. (eds.) Intl. Conf. on
  Computer Aided Verification {(CAV)}. LNCS, vol.~8559, pp. 737--744. Springer
  (2014). \doi{10.1007/978-3-319-08867-9\_49},
  \url{https://doi.org/10.1007/978-3-319-08867-9\_49}

\bibitem{mcsat_arrays}
Dutertre, B., Goel, A., Graham-Lengrand, S., Irfan, A., Jovanovic, D., Mason,
  I.A.: {Yices 2} in {SMT-COMP} 2023  (2023)

\bibitem{GoldwaserMicaliRackoff-SIAM89}
Goldwasser, S., Micali, S., Rackoff, C.: {The Knowledge Complexity of
  Interactive Proof Systems}. SIAM Journal on Computing  \textbf{18}(1),
  186--208 (1989). \doi{10.1137/0218012}

\bibitem{mcsat_bv}
Graham{-}Lengrand, S., Jovanovic, D., Dutertre, B.: Solving bitvectors with
  {MCSAT:} explanations from bits and pieces. In: Peltier, N.,
  Sofronie{-}Stokkermans, V. (eds.) Intl. Joint Conf. on Automated Reasoning
  {(IJCAR)}, Part {I}. LNCS, vol. 12166, pp. 103--121. Springer (2020).
  \doi{10.1007/978-3-030-51074-9\_7},
  \url{https://doi.org/10.1007/978-3-030-51074-9\_7}

\bibitem{hader2023lpar}
Hader, T., Kaufmann, D., Kov{\'{a}}cs, L.: {SMT} solving over finite field
  arithmetic. In: Piskac, R., Voronkov, A. (eds.) Intl. Conf. on Logic for
  Programming, Artificial Intelligence and Reasoning {(LPAR)}. EPiC Series in
  Computing, vol.~94, pp. 238--256. EasyChair (2023). \doi{10.29007/4N6W},
  \url{https://doi.org/10.29007/4n6w}

\bibitem{DBLP:conf/vmcai/Jovanovic17}
Jovanovic, D.: Solving nonlinear integer arithmetic with {MCSAT}. In:
  Bouajjani, A., Monniaux, D. (eds.) Intl. Conf. on Verification, Model
  Checking, and Abstract Interpretation {(VMCAI)}. LNCS, vol. 10145, pp.
  330--346. Springer (2017). \doi{10.1007/978-3-319-52234-0\_18},
  \url{https://doi.org/10.1007/978-3-319-52234-0\_18}

\bibitem{mcsat2}
Jovanovic, D., Barrett, C., de~Moura, L.: The design and implementation of the
  model constructing satisfiability calculus. In: Intl. Conf on Formal Methods
  in Computer-Aided Design {(FMCAD)}. pp. 173--180. IEEE (2013).
  \doi{10.1109/FMCAD.2013.7027033}

\bibitem{jovanovic2017smt}
Jovanovic, D., Dutertre, B.: Libpoly: {A} library for reasoning about
  polynomials. In: Brain, M., Hadarean, L. (eds.) Intl. Workshop on
  Satisfiability Modulo Theories {(SMT)}. {CEUR} Workshop Proceedings,
  vol.~1889, pp. 28--39. CEUR-WS.org (2017),
  \url{https://ceur-ws.org/Vol-1889/paper3.pdf}

\bibitem{nlsat}
Jovanovic, D., de~Moura, L.M.: Solving non-linear arithmetic. In: Gramlich, B.,
  Miller, D., Sattler, U. (eds.) Intl. Joint Conf. on Automated Reasoning
  {(IJCAR)}. LNCS, vol.~7364, pp. 339--354. Springer (2012).
  \doi{10.1007/978-3-642-31365-3\_27},
  \url{https://doi.org/10.1007/978-3-642-31365-3\_27}

\bibitem{mcsat}
de~Moura, L., Jovanovic, D.: A model-constructing satisfiability calculus. In:
  Giacobazzi, R., Berdine, J., Mastroeni, I. (eds.) Intl. Conference on
  Verification, Model Checking, and Abstract Interpretation {(VMCAI)}. LNCS,
  vol.~7737, pp. 1--12. Springer (2013). \doi{10.1007/978-3-642-35873-9\_1},
  \url{https://doi.org/10.1007/978-3-642-35873-9\_1}

\bibitem{DBLP:journals/cacm/MouraB11}
de~Moura, L.M., Bj{\o}rner, N.S.: Satisfiability modulo theories: introduction
  and applications. Commun. {ACM}  \textbf{54}(9),  69--77 (2011)

\bibitem{DBLP:conf/lpar/NieuwenhuisOT04}
Nieuwenhuis, R., Oliveras, A., Tinelli, C.: Abstract {DPLL} and abstract {DPLL}
  modulo theories. In: Baader, F., Voronkov, A. (eds.) Intl. Conf. on Logic for
  Programming, Artificial Intelligence, and Reasoning {(LPAR)}. LNCS,
  vol.~3452, pp. 36--50. Springer (2004). \doi{10.1007/978-3-540-32275-7\_3},
  \url{https://doi.org/10.1007/978-3-540-32275-7\_3}

\bibitem{ozdemir2023cav}
Ozdemir, A., Kremer, G., Tinelli, C., Barrett, C.W.: Satisfiability modulo
  finite fields. In: Intl. Conf. on Computer Aided Verification {(CAV)},
  Part~{II}. LNCS, vol. 13965, pp. 163--186. Springer (2023).
  \doi{10.1007/978-3-031-37703-7\_8},
  \url{https://doi.org/10.1007/978-3-031-37703-7\_8}

\bibitem{ozdemir2023cavzkproofs}
Ozdemir, A., Wahby, R.S., Brown, F., Barrett, C.W.: Bounded verification for
  finite-field-blasting - in a compiler for zero knowledge proofs. In: Enea,
  C., Lal, A. (eds.) Intl. Conf. on Computer Aided Verification {(CAV)},
  Part~{III}. LNCS, vol. 13966, pp. 154--175. Springer (2023).
  \doi{10.1007/978-3-031-37709-9\_8},
  \url{https://doi.org/10.1007/978-3-031-37709-9\_8}

\bibitem{Szabo2018SmartC}
Szabo, N.: {Smart Contracts: Building Blocks for Digital Markets} (1996),
  [Online]. Available: http://www.fon.hum.uva.nl

\bibitem{wang2001elimination}
Wang, D.: {Elimination Methods}. Springer Science \& Business Media (2001)

\end{thebibliography}
\end{document}